\documentclass{article}
\usepackage[utf8]{inputenc}
\usepackage[yyyymmdd]{datetime}
\usepackage{url}
\usepackage{tabularx}
\usepackage{tikz}
\usepackage{wasysym}
\usepackage{flowchart}
\usetikzlibrary{shapes,arrows.meta,chains}

\title{The Forgotten Preconditions for a Well-Functioning Internet}
\author{Geoff Goodell\\Department of Computer Science\\University College London}
{}
\date{\textit{\today}}

\def\smbwd{2cm}

\newcolumntype{L}[1]{>{\raggedright\arraybackslash}p{#1}}
\newcolumntype{C}[1]{>{\centering\arraybackslash}p{#1}}
\newcolumntype{R}[1]{>{\raggedleft\arraybackslash}p{#1}}

\newcommand{\ts}{
    \tikzset{>={Latex[width=3mm,length=3mm]}}
    \tikzstyle{line} = [draw, ->, >=latex, ultra thick]
    \tikzstyle{circ} = [
        draw,
        circle,
        text width=4em,
        text centered,
        minimum width=\smbwd,
        minimum height=0.5cm
    ]
    \tikzstyle{term} = [
        draw,
        terminal,
        text width=4em,
        text centered,
        minimum width=\smbwd,
        minimum height=0.5cm
    ]
    \tikzstyle{box} = [
        draw,
        rectangle,
        align=center,
        text width=8em,
        text centered,
        minimum width=\smbwd,
        minimum height=1cm
    ]
    \tikzstyle{form} = [
        draw,
        trapezium,
        trapezium left angle=70,
        trapezium right angle=-70,
        align=center,
        text width=5em,
        text centered,
        minimum width=\smbwd,
        minimum height=1cm
    ]
    \tikzstyle{wait} = [
        draw,
        trapezium,
        trapezium left angle=-70,
        trapezium right angle=-70,
        align=center,
        text width=5em,
        text centered,
        minimum width=\smbwd,
        minimum height=1cm
    ]
    \tikzstyle{noshape} = [align=center, text centered]
}

\begin{document}

\maketitle

\begin{abstract}

For decades, proponents of the Internet have promised that it would one day
provide a seamless way for everyone in the world to communicate with each
other, without introducing new boundaries, gatekeepers, or power structures.
What happened?  This article explores the system-level characteristics of a key
design feature of the Internet that helped it to achieve widespread adoption,
as well as the system-level implications of certain patterns of use that have
emerged over the years as a result of that feature.  Such patterns include the
system-level acceptance of particular authorities, mechanisms that promote and
enforce the concentration of power, and network effects that implicitly
penalise those who do not comply with decisions taken by privileged actors.  We
provide examples of these patterns and offer some key observations, toward the
development of a general theory of why they emerged despite our best efforts,
and we conclude with some suggestions on how we might mitigate the worst
outcomes and avoid similar experiences in the future.

\end{abstract}

\noindent Policy significance statement: Although the Internet was evidently
designed to be robust and decentralised, some of its design characteristics,
combined with the human tendency to seek protection and convenience, have given
rise to paternalism, exposing control points that enable certain groups to play
outsized roles.  Such groups have managed to effectively corner the market for
four essential aspects of human communication: access, naming, trust, and
reputation.  Registries for names and addresses, certification authorities, and
web infrastructure providers have emerged as de facto gatekeepers, to the
extent that using the Internet in practice implies accepting the primacy of
these actors.  Without a global alternative to the Internet, there is no way to
ensure that the Internet serves the interests of its users.

\section{Introduction}

A preponderance of articles and books have emerged in recent years to decry
pernicious abuse of the infrastructure of the Internet.  Much of the argument
is about \textit{surveillance capitalism}.  We know that companies such as
Facebook undermine human autonomy by creating indexable records of individual
persons and generate tremendous revenue along the way~\cite{zuboff2015}.  We
also know that companies such as Microsoft provide Internet services, such as
Microsoft Teams, to draw ordinary individuals and businesses into walled
gardens that make it possible for the companies to observe their habits and
activities~\cite{waters2021}.  At the same time, we know that some other
businesses have come to depend upon the data harvesting ecosystem, and we worry
that addressing the harms of surveillance capitalism might necessarily entail
collateral damage~\cite{chen2021}.  Economic incentives and entrenched power
dynamics have certainly given rise to the patterns of application service
provision via the software-as-a-service revenue model, and network effects have
buttressed the concentration of resources in the ``cloud'' and data centres.
Arguments about business motivations and economic imperatives are powerful and
moving, and we might hope to mitigate their negative externalities with the
right set of changes to law and regulation, applied carefully.

This view is half right.  But the design of the Internet has a fundamental
property that is less often discussed: \textit{The Internet is designed to
expose network-level metadata about its end-users, both to each other and to
network carriers.} There are important reasons underpinning this design choice,
most notably the fact that Internet users expect the network to carry a packet
to its destination without being told explicitly how to get there.  The guiding
principle has been that the network should ``just work'' with minimal
expectations of knowledge about the structure of the network on the part of end
users and their devices.  This design has many benefits, including both
resilience and dynamism.  If part of the network fails, then the network can
correct itself without the involvement or concern of end-users.  New users,
services, and entire networks can join or leave with minimal impact or cost to
the set of existing participants.  Arguably, this feature was instrumental in
facilitating the early adoption of the Internet.

However, there is a dark side to this design choice as well, which might lead
us to recognise it as a flaw.  As its value in drawing new users waned with the
passage of time, its value as a mechanism of control became apparent.  Exposing
network-level metadata about how end-users are connected to the Internet to
carriers and to other end-users allows for the concentration and abuse of
power, in the form of discrimination, surveillance, and coercion.

The impact of this design choice may have been felt more acutely in recent
years, but the it is certainly not new.  The underlying protocols and
assumptions have been in place for decades, giving rise to control points
related to \textit{access}, \textit{naming}, \textit{trust}, and
\textit{reputation}.  Within each of these categories, the control points have
enabled majoritarian decisions about the locus of authority and penalties for
non-compliance, which have in turn undermined the potential of the Internet to
serve everyone.  Although the design flaw is structural, and even technical, it
offers insight into a foundational problem that is about humanity as much as it
is about technology.

\section{Access: One Internet for all?}

We might say that the Internet is a global institution.  However, this is
misleading.  The Internet is not really institutional, and we might say that
the Internet is successful precisely because it is \textit{not} institutional.
There are no particular rules beyond the protocol specifications and no global
institutions with the power to enforce the few rules that do exist, which
indeed are often broken.  Unlike many global systems, including decentralised
systems such as most cryptocurrency networks, the Internet does not require
global consensus to function, beyond agreement on its foundational protocols.
The Internet has never been globally consistent, and achieving consistency
would be theoretically impossible.  There is no authority that mandates changes
to the protocol, and as of 2021, the world is currently in the process of
migrating the data plane of the Internet from a version originally developed in
1981 to a version originally developed in 1995.  As a ``best effort''
communication medium, the Internet has no mechanisms to ensure quality of
service, or indeed any quality of service at all.  Nonetheless, I shall argue
that all of these characteristics are features, not bugs, in the design of the
Internet, and are essential contributing factors to the success of its design.

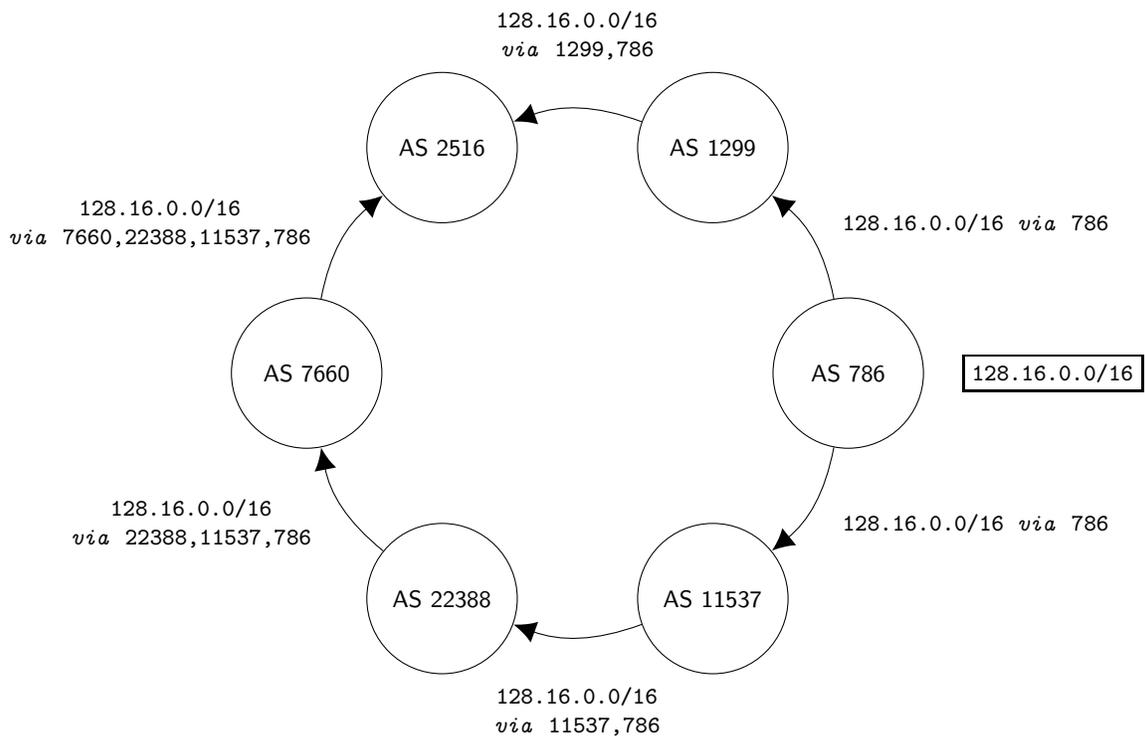
\begin{figure}
\begin{center}
\begin{tikzpicture}[>=latex, node distance=3cm, font={\sf \small}, auto]\ts
\tikzset{>={Latex[width=3mm, length=3mm, line width=0.5mm]}}
\node (as786) at (3.6,0) [circ] {AS 786};
\node (as1299) at (1.8,3) [circ] {AS 1299};
\node (as2516) at (-1.8,3) [circ] {AS 2516};
\node (as11537) at (1.8,-3) [circ] {AS 11537};
\node (as22388) at (-1.8,-3) [circ] {AS 22388};
\node (as7660) at (-3.6,0) [circ] {AS 7660};
\draw[->] (as786) edge[bend right=20] (as1299) {};
\draw[->] (as1299) edge[bend right=20] (as2516) {};
\draw[->] (as786) edge[bend left=20] (as11537) {};
\draw[->] (as11537) edge[bend left=20] (as22388) {};
\draw[->] (as22388) edge[bend left=20] (as7660) {};
\draw[->] (as7660) edge[bend left=20] (as2516) {};
\node (t1) at (5,0) [anchor=west] {\fbox{\texttt{128.16.0.0/16}}};
\node (t2) at (3.4,2) [anchor=west] {\texttt{128.16.0.0/16 \textit{via} 786}};
\node (t3) at (3.4,-2) [anchor=west] {\texttt{128.16.0.0/16 \textit{via} 786}};
\node (t4) at (0,4.5) [align=center] {\texttt{128.16.0.0/16}\\\texttt{\textit{via} 1299,786}};
\node (t5) at (0,-4.5) [align=center] {\texttt{128.16.0.0/16}\\\texttt{\textit{via} 11537,786}};
\node (t3) at (-3.4,-2) [anchor=east,align=center] {\texttt{128.16.0.0/16}\\\texttt{\textit{via} 22388,11537,786}};
\node (t2) at (-3.4,2) [anchor=east,align=center] {\texttt{128.16.0.0/16}\\\texttt{\textit{via} 7660,22388,11537,786}};
\end{tikzpicture}
\end{center}

\caption{\textit{An example of BGP route advertisements.}  AS 786 advertises
prefix \texttt{128.16.0.0/16} to its neighbours, who propagate the
advertisement by appending their respective AS numbers to the path.  AS 2516
has a policy decision to make: Should it send packets destined for
\texttt{128.16.0.0/16} to AS 1299 or to AS 7660?}

\label{f:bgp}
\end{figure}

Every device connected to the Internet speaks Internet Protocol, either the
1981 version or the 1995 version, and has an address (an \textit{IP address}).
The fundamental unit of aggregation in the control plane of the Internet is the
\textit{autonomous system}, or AS.  Autonomous systems communicate with each
other via a protocol called Border Gateway Protocol, or BGP, the first version
of which was developed in 1989.  The purpose of BGP is to exchange reachability
information about autonomous systems~\cite{rfc1105}.  There are over four
hundred thousand autonomous systems, and every IP address that is globally
addressable can be recognised as part of a block (or \textit{prefix}) of
addresses, which in turn is advertised via BGP by an autonomous system.  The
operators of autonomous systems determine, as a matter of local policy, which
advertisements to accept, which not to accept, and which to prioritise over
others (see Figure~\ref{f:bgp}).  Reconciliation is pairwise: if the operators
of an AS receive an advertisement from a neighbouring AS that does not look
right, or if their customers have a problem reaching certain addresses
advertised by a neighbouring AS, then they can raise the matter with their
peers.

There are five regional information registries that manage the allocation of
specific IP address prefixes and autonomous system numbers.  However, these
registries are not really authorities.  Their role is convenience: to prevent
collisions among the numbers that different parties are using, not to
adjudicate the allocation of some scarce resource.  Specific numbers do not
confer any particular advantage, and address prefixes, while limited, are
generally not the factor that limits the ability of an autonomous system to
send and receive traffic.  Furthermore, registries have no ability to enforce
the usage of the numbers in their registries; ultimately, autonomous systems
will route traffic as they please, and allow for the creation of arbitrary
patterns of connection.  So far, so good.

However, to use the Internet, everyone must implicitly agree upon the authority
of this particular set of registrars, since collectively they determine who can
have IP addresses and which addresses they can have.  We all agree, by majority
decision, upon a single globally-meaningful mapping of users to IP addresses.
At the same time, there is no mechanism to ensure that all IP addresses are
equal, and some are more important than others.  In practice, IP addresses are
controlled by gatekeepers.  Control is generally delegated, creating a
hierarchical structure of accountability that can be characterised as a kind of
federation.  A carrier can assign an arbitrary address to a device connected to
a local network, and use \textit{network address translation} to route global
traffic to and from local devices without requiring the local devices to have
globally meaningful names.  As a result, IP addresses have mostly become the
concern of Internet carriers and service providers rather than end-users.  The
ability to differentiate customers on the basis of whether they can be
reachable for new connections allows carriers to offer their services at
different price points, with businesses that maintain servers subsidising
ordinary individuals with personal computers and mobile devices.

Thus, the first manifestation of the structural flaw of the Internet is
exposed: On the basis of reachability alone, the Internet has already been
divided into first-class and second-class citizens.  Unsurprisingly, a primary
motivating factor for the threat to end-to-end system design is revenue for
carriers through price discrimination, not the scarcity of unique addresses,
available bandwidth, or any technical resource in
particular~\cite{odlyzko2004}.  This state of affairs might be seen as an
unintended consequence of a primary incentive for carriers to provide service,
and in fact, many ordinary users with mobile and broadband connections are
indeed assigned globally unique addresses that are purposefully blocked by
their carriers from receiving new connections.

The incentive to create price discrimination, and its concomitant
stratification, exists irrespective of material costs related to infrastructure
development.  The opportunity is available to network carriers as a direct
consequence of the metadata that expose different patterns of use of the
network.  Put another way, network neutrality is foolhardy: Carriers that do
not take advantage of the metadata are leaving revenue on the table and might
even cause their investors to question whether they have disregarded their
fiduciary responsibilities.

For those customers with the good fortune to have Internet devices that are
reachable, the fact that they generally want a way for others to know how to
reach them introduces another risk of unintended consequences, which we shall
explore next.

\section{Naming: Context is everything}

Although the management of AS numbers and IP address prefixes is relatively
peaceful, the management of human-meaningful names for Internet hosts is much
more contentious.  Human-meaningful names are problematic when there is an
expectation that everyone respects the same naming convention.  Who gets to
decide what names we use, and what makes a decision legitimate?  A commonly
recognised principle called Zooko's triangle holds that the names we use cannot
be universally recognised and human-meaningful without being managed by a
common authority~\cite{zooko2001}.  Similar arguments have been made throughout
human history, with memorable parables ranging from the Tower of Babel to
Humpty Dumpty.  Notwithstanding the validity of these longstanding arguments,
the Domain Name System, or DNS~\cite{rfc883}, represents yet another effort to
achieve exactly this kind of global agreement and becomes a second
manifestation of the structural flaw of the Internet.  Perhaps the authors of
the original specification in 1983 had not anticipated the number of Internet
devices that would eventually require names and how important those names would
become.

DNS is hierarchically managed, with a set of globally agreed \textit{root
servers} that delegate authority to a set of \textit{top-level domains}, which
in turn delegate authority to registrars and others who facilitate registration
of names by individuals, institutions, businesses, and other organisations,
usually for a fee.  As with the management of IP addresses, there is a kind of
federation.  However, we all agree, by majority decision, upon the root of the
hierarchy, and all decisions about the binding of names to end-users flow from
that agreement.  The names themselves refer to IP addresses and other metadata
that are used to access Internet services.  A name is said to be \textit{fully
qualified} if it references a specific path of delegation back to the root; an
example of a fully qualified name is \texttt{`www.ntt.com.'}, which indicates
delegation from the root authority to the registrar of \texttt{`com.'} to the
registered authority for \texttt{`ntt.com.'}.  But what organisation has the
privilege to call itself \texttt{ntt.com}?  Short, pithy, and prestigious names
are scarce, and in practice they demand a hefty premium in the global
marketplace for names, if they are traded at all.

To make matters worse, the use of DNS generally involves the active involvement
of another gatekeeper, the client-side resolver.  The client-side resolver can
observe the hostnames requested by the client, potentially exposing the client
to profiling and discrimination.  Over the years, various approaches such as
DNSSEC~\cite{rfc2065} and DNS over TLS~\cite{rfc7858} have been developed to
reduce the risk that a client's carrier might manipulate the names that a
client receives.  However, these approaches are not really private.  While they
protect the data inside the DNS requests from eavesdropping by client-side
carriers, they shift the locus of eavesdropping to global-scale platform
service providers instead, such as Google or Cloudflare.  Such platform service
providers amass much more data than most local carriers ever could, benefiting
their own profiling activities and potentially attracting attacks.  At the same
time, the local network can still determine the sites clients visit, since the
clients generally send traffic to the addresses they resolve.

Ostensibly in response to criticism, Cloudflare recently partnered with Apple
and Fastly to propose ``Oblivious DNS over HTTPS'' (ODoH), which adds
end-to-end encryption and a proxy service to stand between DNS clients and its
DNS resolvers~\cite{singanamalla2021}.  However, since the proxy service and,
presumably, its network carriers learn both the IP address of the client as
well as the exact timing and sequence information of its contact with the
resolver, this single-hop proxy does not offer much privacy.  In particular,
Cloudflare acknowledges relationships with ``leading proxy partners, including
PCCW Global, SURF, and Equinix''~\cite{cloudflare-odoh}, and both parties could
easily collaborate to de-anonymise a client, if not as part of routine
business, then certainly when compelled by an authority to do so.  Even if it
were expected that clients would mainly communicate with the resolver via
independent anonymising proxy networks, operators of large-scale DNS resolvers
would still be positioned to exercise outsize authority over Internet names.

Inexorably, the globally recognised allocation of names to registrants
introduces a power dynamic and raises the question of fairness.  After all, why
are some parties privileged to the exclusion of others?  There are various
initiatives, such as Namecoin~\cite{namecoin}, that seek to reject the
authority of DNS entirely and replace it with a transparent marketplace, but
such attempts ultimately introduce more problems than they solve.  For example,
can any system that is first-come, first-served can ever be fair?  Should the
best name always go to the highest bidder?  It would seem that both the
paternalistic hand of a trusted authority and the invisible hand of the
marketplace fall short of furnishing a solution that works for everyone,
precisely because any solution that demands global consensus from a process
must also assume global and durable acceptance of the outcome of that process.

A better approach to naming would be to recognise that in human settings, names
are not bestowed by authorities except for specific contexts in which the
authorities exercise their authority.  In general, individual persons and the
local communities in which they are embedded decide which names to use to
identify ordinary people, groups, objects, concepts, and so on.  Perhaps we
might encourage users to do what humans have always done, and assign
\textit{their own} names for the parties and services with whom they interact.
An approach described as ``Petnames''~\cite{petnames} can facilitate this.  A
major selling point of hyperlinks is the idea that users can navigate without
explicitly concerning themselves with the official universal resource
identifiers (URIs) of the resources they seek.  Perhaps users should assign
their own names using tools akin to browser bookmarks, and the sites themselves
could even suggest appropriate names to use.  Freeing universally recognisable
URIs from the requirement to be human-meaningful allows for the possibility
that they might be made secure.

\section{Trust: Anchors or millstones?}

Since Internet routing is fundamentally decentralised, how can one be sure that
the other party to a conversation, as identified by its IP address or domain
name, is authentic?  This is a question of security, and we know that we can
use cryptography not only to protect our conversations from eavesdropping but
also to verify the authenticity of a conversation endpoint using digital
certificates~\cite{rfc2459}.  This is straightforward if we can directly and
securely exchange public keys with everyone that we might want to engage in a
conversation via the Internet.  In practice, however, people use the Internet
to engage with many different kinds of actors and seldom exchange certificates
directly.  When was the last time your bank advised you to verify the
fingerprint of its public key?  Thus, we have identified a third structural
flaw of the Internet: By providing for the establishment of regional or even
global authorities, its design has bypassed local institutions and authorities
as mechanisms for creating trust, and a popular shortcut, which we describe in
this section, has undermined those mechanisms and institutions.  Although
decentralised governance is recognised as promoting ``greater accountability
and better services'', it also tends to deliver less ``technical and governance
capacity''~\cite{smoke2015}.  Individuals and local communities are in no
position to negotiate with global authorities, but has the Internet raised the
bar for technical and governance capacity beyond what local authorities can
manage?

Most modern operating systems and web browsers come pre-installed with a set of
so-called \textit{trust anchors} that serve as trusted third parties to verify
the public keys associated with particular IP addresses or domain names.  In
principle, users could add or remove trust anchors from this list according to
their personal preferences, but almost nobody actually does.  Moreover, since
Internet services must present their certificates to web browsers and other
client software, the question facing the operators of those services is: Which
trust anchors are the web browsers and other applications running on end-user
devices likely to accept?  The operators then seek signatures from the trust
anchors that are commonly shipped with end-user software and present those
signatures in the form of certificates.  Since obtaining, storing, and
presenting certificates from certificate authorities carry operational costs
(and sometimes economic costs, although the emergence of free
alternatives~\cite{letsencrypt} has changed this), the operators are strongly
motivated to be parsimonious.  Thus, we have an implicit agreement between site
operators and software distributors about the correct set of trust anchors to
use, and as a result, those trust anchors become powerful gatekeepers.  Just as
with access and naming, and consistently with the pattern of majoritarianism,
end-users have been forced to accept what has already been agreed.

What happens when a trust anchor fails?  Routine breaches of privileged
information such as private keys take place from time to time and do not
constitute a theoretical question.  For example, consider the well-publicised
compromise of Comodo in March 2011~\cite{comodo}.  Because certificate
authorities are trusted by default by widespread client software, the stakes of
a breach are high.  The compromise of private keys of certificate authorities
enabled the Stuxnet worm, which was discovered in 2010 and had been used in
attacks against an Iranian nuclear facility~\cite{anderson2012}, and a
subsequent, successful attack on DigiNotar, another certificate authority,
allowed the interception of Internet services provided by
Google~\cite{adkins2011}.  If vesting so much power in a small number of
globally trusted institutions and businesses seems like a dangerous idea,
that's because it is.

One initiative, Certificate Transparency~\cite{rfc9162}, aims to address the
security problem by creating audit logs to publicise known certificates issued
by certification authorities.  However, because its primary effect is to shift
the locus of control from the certification authorities themselves to the
maintainers of the audit logs, it is not really a satisfying answer to the
question: \textit{Quis custodiet ipsos custodes?}  By implicitly anticipating
that all certificates would be logged publicly, it also raises the question of
whether it should be possible for legitimate trust relationships to be
established and maintained outside the public view.  If the Tower of Babel
teaches us that the public cannot agree on the authority of one set of trust
anchors, then why would the public believe in the ultimate authority of one
public logging facility and its ecosystem of maintainers and users?


\section{Reputation: Credit or extortion?}

DNS is a formal system with formal rules and institutionally trusted
authorities, and the secure verification of Internet services also carries the
weight of trusted institutions.  However, just as not all power is
institutional, not all structural shortcomings of the Internet can be
characterised as concerns about the weaknesses and illegitimacy of trusted
authorities.  A mafia family might be expected to serve the interests of a
local community by providing protection, and whether such protection is
legitimate might be a matter of perspective, perhaps even ethically debatable.
Arguably, such behaviour is endemic to the Internet, particularly given its
common role as a venue for anti-social and threatening behaviour.  The
mechanisms that enable such behaviour to emerge and flourish represent a fourth
structural flaw of the Internet.

Scale is a powerful tool for creating change in the world.  However, only a
handful of actors can wield it, and those who can usually have an incentive not
to change the environment that enabled them to thrive.  The counterexamples,
however, illustrate both the opportunities and the danger.

For context, consider how a powerful organisation, Google, used its scale to
change the Internet.  In 2012, a team at Google sought to optimise the
performance of communication between its servers, which were increasingly
carrying multimedia content, and personal computers running its Chrome web
browser; the team designed and implemented a new transport protocol called
QUIC~\cite{roskind2012}.  Because Google controlled both a significant volume
of web servers (through its online services) and a significant share of web
browsers (through Chrome), this protocol quickly reached broad adoption and
experienced a relatively swift journey through the standardisation
process~\cite{hamilton2016,rfc9000}.  The protocol is widely used today.
Adoption of the protocol was generally voluntary, and so this use of power can
be considered benign.

When scale is combined with the problem of assessing creditworthiness or
reputation, the results are somewhat less benign.  Consider the following case
that again involves Google and relates to e-mail spam.  No one really likes to
receive unsolicited messages from dodgy actors, which is a system-level
consequence of the common practice of using the same e-mail address across
different contexts, a topic for a separate article.  A technical approach
to mitigating spam is to require senders to prove that they are the rightful
owners of the domain names they use to represent themselves; this approach
forms the essence of the Sender Policy Framework, or SPF~\cite{rfc4408}, which
was first proposed in 2000~\cite{dmarcian}, as well as DomainKeys Identified
Mail, or DKIM~\cite{rfc4871}, which was introduced in 2007.  Notice that this
approach further entrenches the authority of DNS system operators.  One might
imagine that it would be impossible to compel all of the mail servers on the
Internet to stop sending mail without valid SPF or DKIM signatures, and indeed
both the specification for SPF and the specification for DKIM advise mail
servers not to reject messages `solely' on the basis of an inability to
authenticate the sender~\cite{rfc4408,rfc4871}.  However, one mail server
operator, Google, implemented a policy that stretched the limits of this
recommendation by routinely classifying mail sent to users of its popular Gmail
service that did not include a valid SPF or DKIM header as
spam~\cite{google2017}.  As a result, many mail servers were forced to
implement SPF or DKIM because their users wanted to be able to send mail to
Gmail users.  And so, by leveraging its scale, Google had successfully managed
to twist the arms of mail server operators around the world.

A decidedly less benign example involves the creation of blacklists of IP
addresses on the basis of their reputation, ostensibly for the purpose of
mitigating attacks on servers.  The idea of using routing information, such as
IP address, as a convenient way to judge the reputation of a sender is not
new~\cite{goodell2007}, and the ability to wield disproportionate power is not
always limited to large, wealthy corporations.  Consider the case of SORBS, an
Australian organisation that aims to reduce the preponderance of spam through
the publication of a list of IP addresses associated with spam or abuse.
Although this site does not directly engage in filtering traffic, many e-mail
server operators configure their servers to consult this list as part of their
routine operation and flag messages as spam or reject them outright if they are
received from a mail server whose IP address appears on the SORBS list.  SORBS
generally requires the operators of servers with blacklisted IP addresses to
explicitly request expungment, subject to the discretion of SORBS
staff~\cite{sorbs2021}, and for years, such operators were required to give
money to charity as well~\cite{sorbs2010}.  Similarly, services such as
Spamhaus maintain lists of IP addresses that are apparently ``directly involved
in malicious behaviour'' or have ``a bad reputation associated with
them''~\cite{spamhaus}.  Carriers are incentivised to maintain discipline among
their subscribers, perhaps even to threaten them with termination, if such
services list their IP addresses.

A related example involves the practice of using similar blacklists to restrict
access to websites.  For example, the popular web infrastructure provider
Cloudflare~\cite{cloudflare} offers its customers an option to block IP
addresses associated with public VPNs and anonymising proxy networks such as
Tor~\cite{tor}.  In 2015, Akamai Technologies released a report claiming that
in an experiment conducted by its researchers, ``Tor exit nodes were much more
likely to contain malicious requests'' and recommended that traffic from Tor
should be either heavily scrutinized... or completely blocked'' using tools
such as its proprietary security module ``Akamai Kona Site
Defender''~\cite{akamai2015}.  Kona Site Defender includes a service called
``Client Reputation'' that categorises IP addresses as being malicious or not
on the basis of their interaction with Akamai services around the
world~\cite{akamai,kona}, which its customers can use to block access to their
websites.

A direct system-level consequence of this pattern of blacklisting is that
Internet users who share IP addresses with each other, and especially Internet
users who rely upon anonymity networks for their privacy, are denied access to
much of the Internet.  A second-order system-level consequence is that Internet
users are forced to relinquish their privacy and must instead consider
maintaining favourable reputations in the eyes of powerful gatekeepers, as well
as strong relationships with their Internet carriers, as a prerequisite to
using the Internet.


In all such cases, extortionate behaviour is enabled by the separation of the
function of maintaining the blacklist from the choice to use the blacklist, the
perceived benefit of the blacklist on the part of its users, and the lack of
accountability with respect to policy on the part of server operators that use
the blacklist to the users of their own services.  Are these the stewards that
we wanted?  Certainly they are the stewards that we have.


\section{Negative externalities}

All of our distributed systems, however good they are, are susceptible to abuse
of the power that has been vested in the systems themselves.  In particular,
systems based on the Internet reflect, at minimum, the unfairness and power
dynamics that are intrinsic to the Internet itself.

This is a tragedy of the commons.  The same control points that are called upon
to attract users by offering convenience are inevitably exploited to control
them.  The features that allow key users and service providers to privately
benefit introduce significant public costs.  It is easy to see how those with
an inclination to ignore the effects of externalities in mechanism design might
be attracted to the features of the Internet, wherein there are ample
opportunities to profit by imposing diffuse, often imperceptible costs upon a
broad set of users.

An important lesson of the Internet is that users mostly do not understand what
can be taken away from them until after it is gone.  As metadata leakage
furnishes control to powerful gatekeepers, the rest of the world shrugs, having
no recourse.  Having achieved tremendous power and influence, platforms defy
efforts to resist their locus of control.  It might seem that individuals have
a choice to ignore the power of platforms, but they do not: The value of a
platform is defined by its network effects~\cite{baron2022}.  This is true not
only for popular applications and social media platforms, but also for the
implicit structures under the surface that define the rules for routing,
naming, public access, and public trust.

\section{Conclusion: Protecting our infrastructure from ourselves}

It is worth considering the original design of the Internet as a reminder that
it is possible to build a system that avoids vesting too much power in in its
operators.  This concept is generally described as the \textit{end-to-end
principle} and achieved widespread appreciation well before the protocols
described earlier in this article were developed~\cite{saltzer1984}.  We find
that the Internet is not majoritarian per se, but weaknesses in its design have
allowed certain actors to develop and exercise power over others.  But, how do
we undo the dangerous power structures that have become part of the de facto
architecture of the Internet?  Moreover, how do we ensure that the systems we
design in the future do not become centralised in the first place?

We note that global consensus and, more generally, the interest in establishing
an authoritative global view of something, is exactly where we see the greatest
risk of centralisation in practice.  Part of the interest in an authoritative
global view derives from the convenience of not having to worry about whether
we share context with our neighbours, although it is perhaps when there is a
perception of risk that the tendency to embrace an authoritative global view is
most pronounced.  Under such circumstances, we collectively retreat to the
safety of big, dominant actors that can exercise their control and, in so
doing, help us avoid the responsibility of decision-making and the potential
blame that might follow~\cite{jackall2009}.  Our observation should be a
warning sign for those who seek to build systems that rely upon global
consensus.  Do systems that require global  consensus grind to a halt when
anyone disagrees, or do they implicitly ignore the wishes of the minority?  Put
another way, who or what is the arbiter, and what is to prevent that arbiter
from being compromised by those who disagree from us?

With respect to the Internet, there is no easy solution, although we can
imagine its characteristics.  Identity must be non-hierarchical and separate
from routing, so that reputations are not earned and lost based upon how users
are connected to the network.  Internet routers must not be able to infer who
is speaking to whom, or what kid of information is being communicated, from the
traffic they carry.  This would seem to imply a requirement for a second,
\textit{oblivious} layer of routing between the existing network layer that
relies upon convenient connectivity and the end-to-end protocols that users
see.  Perhaps this could be achieved by encouraging and expecting Internet
users to use onion routing~\cite{syverson1996,tor} as a way to shield
end-to-end conversations from the pastoral gaze of network operators, and to
use self-certifying names, such as those used by Tor onion services, to prevent
the accumulation and exercise of illegitimate authority by third-party actors.
Self-certifying names should be opaque, to reduce the likelihood that people
would confuse them with meaningful names or concepts.  It is not far-fetched to
imagine inserting another layer to the Internet, between the layer at which
routing operates and the layer that manages end-to-end conversations, to
mitigate the harm associated with authorities occupying positions of control
within the network.

Finally, it is in the interest of Internet users to avoid being pinned to a
single identity, either over time or across their many relationships, and they
should have tools that help them assume different identities for their
different relationships, to mitigate the risk of both spam and blacklisting at
the same time.

Above all, we should eliminate the assumption that metadata exposed to the
system will not be misused, as well as the assumption of paternalistic goodwill
on the part of powerful actors and service providers.  It would behoove us to
consider the fundamental trade-off intrinsic to all global systems and pursue a
principle of parsimony: How can we provide the system with enough power to do
what we need it to do, without providing it with enough power to profit at our
expense?  We must take a clear-eyed view to decisions that were made long ago
in the interest of marshalling attention and serving the narrow interests of
actors with a role in developing key infrastructure, but which have since
outlasted their usefulness.  It is time to rethink the relationship between the
network and metadata: The Internet is not fully mature until the edges can
communicate without the permission of the centre.

\section*{Acknowledgements}

We express gratitude for the feedback of anonymous reviewers and attendees of
the CoGMa Workshop, as well as the continued support of Professor Tomaso Aste,
the Centre for Blockchain Technologies at University College London, and the
Systemic Risk Centre at the London School of Economics.  A pre-print of this
article is also available online~\cite{arxiv-2109}.

\noindent Competing interest statement: The author declares no competing
interests.

\noindent Funding statement: The author acknowledges the support of TODAQ
Financial and the Peer Social Foundation.

\noindent Data availability statement: Not applicable.

\noindent Author contributions: Writing, review, and editing: G.G.

\end{document}